\magnification=1200
\null\bigskip\bigskip
\centerline{EXTENDING THE DEFINITION OF ENTROPY}
\medskip
\centerline{TO NONEQUILIBRIUM STEADY STATES.}
\bigskip
\centerline{by David Ruelle\footnote{*}{IHES (91440 Bures sur Yvette,
France) $<$ruelle@ihes.fr$>$, and Math. Dept., Rutgers University
(Piscataway, NJ 08854-8019, USA).}}
\bigskip\noindent
{\bf Abstract.}  {\sl We study the nonequilibrium statistical mechanics of a finite classical system subjected to nongradient forces $\xi$ and maintained at fixed kinetic energy (Hoover-Evans isokinetic thermostat).  We assume that the microscopic dynamics is sufficiently chaotic (Gallavotti-Cohen chaotic hypothesis) and that there is a natural nonequilibrium steady state $\rho_\xi$.  When $\xi$ is replaced by $\xi+\delta\xi$ one can compute the change $\delta\rho$ of $\rho_\xi$ (linear response) and define an entropy change $\delta S$ based on energy considerations.  When $\xi$ is varied around a loop, the total change of $S$ need not vanish: outside of equilibrium the entropy has curvature.  But at equilibrium (i.e. if $\xi$ is a gradient) we show that the curvature is zero, and that the entropy $S(\xi+\delta\xi)$ near equilibrium is well defined to second order in $\delta\xi$.}
\bigskip\noindent
{\bf Introduction.}
\medskip
	The purpose of the present paper is to discuss the statistical mechanics of a finite physical system maintained in a nonequilibrium steady state at a constant temperature.  In such a system, entropy is produced at some constant rate $\ge0$.  Here we investigate the possibility of also associating a finite entropy $S$ with our nonequilibrium system, extending the definition of equilibrium entropy.  We restrict our discussion to the case of a classical system with an {\it isokinetic thermostat} (as defined by W. Hoover [9] and D. Evans [6], see below).  
\medskip
	If $\rho(dx)=g(x)dx$ is the probability measure in phase space corresponding to an equilibrium state, the corresponding {\it Gibbs entropy} is 
$$	S(\rho)=-\int dx\,g(x)\log g(x)      $$
The probability measure $\rho(dx)$ describing a nonequilibrium steady state is in general singular with respect to $dx$, and the corresponding Gibbs entropy is thus $-\infty$.  To extend the definition of entropy outside of equilibrium we shall use another idea, based on the thermodynamic relation $\delta S=\delta Q/T$, where $\delta Q$ is energy exchanged, and $T$ the absolute temperature.
\medskip
	We consider a finite mechanical system in a nonequilibrium (in general) steady state $\rho_\xi$ under the effect of a nongradient (in general) force $\xi$ and an isokinetic thermostat at temperature $\beta^{-1}$.  We give below a definition of the entropy increment $S(\xi\to\xi+\delta\xi)$ corresponding to a small increment $\delta\xi$ of $\xi$.  Our definition is based on energy exchanged, uses the microscopic dynamics of the system, and agrees with the equilibrium statistical mechanics definition when $\xi$ and $\delta\xi$ are gradient forces, {\it i.e.}, for equilibrium situations.  Outside of equilibrium, for a loop $\xi\to\ldots\to\xi$, the sum $S({\rm loop})$ of the entropy increments in not expected to vanish in general.  This means that the ``entropy connection'' has a curvature.  Since $S({\rm loop})$ is of second order in the size of the loop, the increment $S(\xi\to\xi+\delta\xi)$ is well defined to first order in $\delta\xi$.  If $\xi$ is a gradient the curvature vanishes and therefore the entropy close to equilibrium
$$	S(\xi+\delta\xi)=S(\xi)+S(\xi\to\xi+\delta\xi)      $$
is well defined to second order in $\delta\xi$.  
\medskip
	Systems outside of equilibrium exhibit a variety of phenomena, like metastability and hysteresis, which we want to exclude here.  We assume that a nonequilibrium steady state is naturally defined, and we study its variations under parameter changes by using the techniques of the ergodic theory of differentiable dynamical systems.  Basically we assume that the microscopic dynamics is sufficiently {\it chaotic} (this is the content of the ``chaotic hypothesis'' of G. Gallavotti and E. Cohen [8]).  Our nonequilibrium steady state $\rho_\xi$ is then a {\it natural} or SRB {\it  measure} and we apply linear response theory (D. Ruelle [13]) to determine changes of $\rho_\xi$ for variations $\xi\to\xi+\delta\xi$.  The linear response is given by integrals over time which generalize those appearing in the fluctuation-dissipation theorem.  
\medskip
	Mathematical proofs of the linear response formulas are within reach under suitable assumptions of uniform hyperbolicity.  (A hyperbolic system with singularities and isokinetic thermostat close to equilibrium has been rigorously studied in [2]).  But, in general, uniform hyperbolicity assumptions are unrealistically strong from a physical point of view.  The present paper is thus meant as theoretical physics rather than mathematical physics.  There is a leap of faith in believing that our linear response formulas apply to any given physical setup, but the situation is not worse than for applications of the fluctuation-dissipation theorem.  
\medskip
	In [14] another approach to the definition of entropy outside of equilibrium was proposed (Lyapunov entropy) replacing phase space volume by volume in a suitable (Kaplan-Yorke) reduced dimension.  This idea was taken up in [7] where an attempt is made at defining the entropy in the large system limit.  We do not investigate here this limit.  Physically, the entropy is best defined in the large system limit as the {\it Boltzmann entropy}, a concept based on the phase space volume associated with a given macrostate, and vigorously defended by J. Lebowitz [11].  It remains to be verified if the definition of entropy given in the present paper can be related to the Boltzmann entropy.  One would also like to check that our results are not tied to the use of the isokinetic thermostat, but extend to more general situations.  (The isokinetic thermostat is very convenient for calculations, but does not quite reproduce the Hamiltonian time evolution at equilibrium).  
\medskip\noindent
{\bf Isokinetic time evolution.}
\medskip
	We consider the classical time evolution 
$$ {d\over dt}\pmatrix{p\cr q\cr}=\pmatrix{\xi-\alpha p\cr p/m\cr}\eqno{(1)} $$
where $p,q\in{\bf R}^N$.  We shall also use the notation $x=\pmatrix{p\cr q\cr}\in{\bf R}^{2N}$ and rewrite (1) as 
$$	{dx\over dt}=F_\xi(x)\eqno{(2)}      $$
The Euclidean scalar product of vectors $a$, $b$ in ${\bf R}^N$ or ${\bf R}^{2N}$ will be denoted by $a\cdot b$.  The force $\xi=\xi(q)$ is not necessarily a gradient, and we take 
$$	\alpha=\alpha(x)={p\cdot\xi(q)\over p\cdot p}      $$
so that 
$$	{d\over dt}\big({p\cdot p\over 2m}\big)=0      $$
The term $-\alpha p$ in (1) corresponds to the much discussed {\it isokinetic thermostat} (a special case of the Gaussian thermostat of Hoover and Evans, see [9], [6], [10]).  We shall denote by $(f_\xi^t)_{t\in{\bf R}}$ the flow defined by (2), {\it i.e.} $f_\xi^tx$ is the solution at time $t$ corresponding to the initial condition $x$.  
\medskip\noindent
{\bf Entropy changes.}
\medskip
	The local rate $e_\xi(x)$ of volume contraction corresponding to the vector field $F_\xi$ is minus its divergence, and easily computed to be $\Phi(x)=(N-1)\alpha(x)$.  This is identified with the local rate of entropy production (see [1], [3]).  When integrated over a nonequilibrium steady state $\rho_\xi(dx)$, it gives the corresponding global rate of entropy production.  It is natural to define the change of entropy $S(\xi\to\xi+\delta\xi)$ to be the entropy released in the time interval $[0,+\infty)$ when the force $\xi+\delta\xi$ acting during the interval $(-\infty,0)$ is replaced by $\xi$ in the interval $[0,+\infty)$.   At time $t\ge0$ our system is in a state $\rho_\xi+\delta_t\rho$ which reduces to $\rho_{\xi+\delta\xi}$ at $t=0$ and tends to $\rho_\xi$ when $t\to\infty$ (an expression for $\delta_t\rho$ will be given below).  We have thus to first order in $\delta\xi$
$$	S(\xi\to\xi+\delta\xi)
	=\int_0^\infty dt\int\delta_t\rho(dx)\,e_\xi(x)\eqno{(3)}      $$
\noindent
{\bf Dynamical assumptions.}
\medskip
	In order to proceed we need now to make some assumptions on the dynamics defined by (1) and on the measure $\rho_\xi$.  As we have said, we want the time evolution to be sufficiently chaotic, {\it i.e.}, the flow $(f_\xi^t)$ to be hyperbolic in some mathematical sense, and the nonequilibrium steady state $\rho_\xi$ to be an SRB measure.  For our purposes we can define an SRB measure as a limit $\lim_{t\to+\infty}(f_\xi^t)^*\sigma$ where $\sigma$ is absolutely continuous with respect to $dp\,dq$ conditioned to $\{(p,q):p\cdot p/2m=K\}$.  An SRB measure is usually singular, but ``smooth along unstable directions''.  For a physical discussion of the present setup see [14].  We shall also assume exponential decay of correlations (see [4], [5], [12]).  As a consequence of these assumptions we have the following linear response formula (see [13]): 
$$	\delta_t\rho(\phi)=\int_{-\infty}^td\tau\,\rho_\xi
\big(\nabla_x(\phi\circ f_\xi^{t-\tau})\cdot\delta_\tau F(x)\big)\eqno{(4)}  $$
where $\delta_tF$ is a time dependent small perturbation of the right-hand side $F_\xi$ of (2), and $\delta_t\rho$ is the corresponding perturbation of $\rho_\xi$ at time $t$.  The integral over $\tau$ converges exponentially.  The test function $\phi$ is assumed to be differentiable because $\delta_t\rho$ is in general a distribution rather than a measure.  We have written $\rho_\xi(\phi)=\int\rho_\xi(dx)\phi(x)$ and similarly for $\delta\rho$.  Note that for time independent $\delta F$ the time independent $\delta\rho$ is given by 
$$	\delta\rho(\phi)=\int_0^\infty ds\int\rho_\xi(dx)
	\nabla_x(\phi\circ f_\xi^s)\cdot\delta F(x)      $$
\noindent
{\bf Notation.}  
\medskip
	We have defined
$$	F_\xi(x)=\pmatrix{\xi-\alpha p\cr p/m\cr}\quad,
	\quad\alpha=\alpha(x)={p\cdot\xi(q)\over p\cdot p}\quad,
	\quad e_\xi(x)=\Phi(x)=(N-1)\alpha(x)\eqno{(5)}      $$
We shall use infinitesimal perturbations $\delta\xi=\xi^i$, $\delta F=F^i$ (no time dependence, $i=1,2$) and let
$$	\matrix{F^i=\pmatrix{\xi^i-\alpha^i p\cr 0\cr}\quad,
	\quad G=\pmatrix{0\cr\xi\cr}\quad,
	\quad G^i=\pmatrix{0\cr\xi^i\cr}\cr
	\alpha^i=p\cdot\xi^i/p\cdot p\quad,
	\quad\Phi^i(x)=(N-1)\alpha^i}\eqno{(6)}      $$
We also denote by $K$ the kinetic energy (conserved by (1)), and let $\beta^{-1}$ be the corresponding temperature:
$$	K={p\cdot p\over2m}\qquad,\qquad\beta={N-1\over2K}      $$
We shall from now on write $f_\xi^t=f^t$ and $\rho_\xi=\rho$.
\medskip\noindent
{\bf Proposition.}
\medskip
	{\sl With the above notation let 
$$	\gamma_\xi(\xi^1)=\int_0^\infty ds\int_0^\infty dt
\int\rho(dx)\,\nabla_x(\Phi\circ f^{s+t})\cdot F^1(x)\eqno{(7)}      $$
define a linear form in $\xi^1$.  Then, to first order in $\xi^1$,  
$$	S(\xi\to\xi+\xi^1)=\gamma_\xi(\xi^1)\eqno{(8)}      $$}
\indent
	Using (3),(5),(4),(6) we have indeed
$$	S(\xi\to\xi+\xi^1)=\int_0^\infty dt\int\delta_t\rho(dx)\Phi(x)
	=\int_0^\infty dt\int_{-\infty}^0d\tau
	\rho(\nabla_x(\Phi\circ f^{t-\tau})\cdot F^1(x))      $$
and replacing $t$ by $s$, $\tau$ by $-t$ gives (8).  
\medskip\noindent
{\bf Curvature.}
\medskip
	To second order in $\xi^1$ we have 
$$	S(\xi\to\xi+\xi^1)=\int_0^1d\lambda\,\gamma_{\xi+\lambda\xi^1}(\xi^1)
	=\gamma_\xi(\xi^1)+{1\over2}(D_\xi\gamma_\cdot(\xi^1))(\xi^1)      $$
where $D_\xi\gamma_\cdot(\xi^1)$ is the functional derivative of $\gamma_\xi(\xi^1)$ with respect to $\xi$.  And an easy second order calculation gives 
$$	S(\xi\to\xi+\xi^1)+S(\xi+\xi^1\to\xi+\xi^1+\xi^2)
	+S(\xi+\xi^1+\xi^2\to\xi+\xi^2)+S(\xi+\xi^2\to\xi)      $$
$$	=R_\xi(\xi^1,\xi^2)      $$
where the curvature form $R_\xi$ is defined by 
$$	R_\xi(\xi^1,\xi^2)
	=(D_\xi\gamma_\cdot(\xi^2))(\xi^1)
	-(D_\xi\gamma_\cdot(\xi^1))(\xi^2)\eqno{(9)}      $$
If $C$ is a closed curve in the space of force fields $\xi$, the change of entropy corresponding to turning around the curve is $\oint_C\gamma_\xi(d\xi)$.  It is of second order in the size of the curve if $R_\xi\ne0$, of higher order if the curvature vanishes.  
\medskip\noindent
{\bf Proposition.}
\medskip
	{\sl Define a bilinear form in $\xi^1$, $\xi^2$ by 
$$	\gamma_\xi(\xi^1,\xi^2)=\int_0^\infty ds\int_0^\infty dt
\int\rho(dx)\,\nabla_x(\Phi^1\circ f^{s+t})\cdot F^2(x)\eqno{(10)}      $$
Assume now that $\tilde\xi$ is locally gradient and write $\tilde G=\pmatrix{0\cr\tilde\xi\cr}$.  Then
\medskip
	(i) $\gamma_\xi(\xi^1)=\gamma_\xi(\xi,\xi^1)$
\medskip
	(ii) $\gamma_\xi(\tilde\xi,\xi^1)=-\beta\int_0^\infty ds\int\rho(dx)\,\tilde G(f^sx)\cdot(T_xf^s)F^1(x)$
\medskip
	(iii) $(D_\xi\gamma_\cdot(\tilde\xi,\xi^1)(\xi^2)$
$$	=-\beta\int_0^\infty ds\int_0^\infty dt\int\rho(dx)\,
	[\nabla_x(\tilde\Psi^{1s}\circ f^t)\cdot F^2(x)
	+\nabla_x(\tilde\Psi^{2s}\circ f^t)\cdot F^1(x)]      $$
where $\tilde\Psi^{is}(x)=\tilde G(f^sx)\cdot(T_xf^s)F^i(x)$.  
\medskip
	(iv) $(D_{\xi}\gamma_\eta(\cdot,\xi^1))(\xi^2)=\gamma_\eta(\xi^2,\xi^1)$
\medskip
	(v) If $\xi$ is locally gradient, then }
$$  R_\xi(\xi^1,\xi^2)=\gamma_\xi(\xi^1,\xi^2)-\gamma_\xi(\xi^2,\xi^1)  $$ 
\indent
	(i) follows directly from the definitions (7) and (10).  
\medskip
	The assuption that $\tilde\xi$ is locally gradient means that we have a configuration space $D\subset{\bf R}^N$ which is not simply connected and that $\tilde\xi(q)=-\nabla_q\tilde V$ where $\tilde V$ is a ``multivalued function'' on $D$.  Writing $f^tx=\pmatrix{p(t)\cr q(t)\cr}$, $\tilde\Phi=(N-1)p\cdot\tilde\xi/p\cdot p$ we have 
$$\int_0^\infty dt\,\tilde\Phi\circ f^{s+t}=m{N-1\over p\cdot p}
	\int_0^\infty dt\,{d\over dt}q(s+t)\cdot\tilde\xi(q(s+t))      $$
$$    =\beta\lim_{T\to\infty}[\tilde V(q(s))-\tilde V(q(s+T))]    $$
hence 
$$	\gamma_\xi(\tilde\xi,\xi^1)=\beta\lim_{T\to\infty}\int_0^\infty ds
\int\rho(dx)[\nabla_x\tilde V(q(s))-\nabla_x\tilde V(q(s+T))]\cdot F^1(x)    $$
$$	=\beta\lim_{T\to\infty}\int_0^T ds
	\int\rho(dx)\nabla_x\tilde V(q(s))\cdot F^1(x)
	=\beta\int_0^\infty ds\int\rho(dx)\pmatrix{0\cr\nabla_{q(s)}\tilde V}
	\cdot(T_xf^s)F^1(x)      $$
$$=-\beta\int_0^\infty ds\int\rho(dx)\,\tilde G(f^sx)\cdot(T_xf^s)F^1(x) $$
which proves (ii).  
\medskip
	We have thus
$$	\gamma_\xi(\tilde\xi,\xi^1)
	=-\beta\int_0^\infty ds\int\rho(dx)\,\tilde\Psi^{1s}(x)$$
$$	\tilde\Psi^{1s}(x)=\tilde G(f^sx)\cdot(T_xf^s)F^1(x)
	=-\nabla_x\tilde V(q(s))\cdot F^1(x)      $$
Therefore, since $\rho$ and $f^s$ both depend on $\xi$,
$$	-D_\xi(\gamma_\cdot(\tilde\xi,\xi^1))(\xi^2)
=D_\xi[\beta\int_0^\infty ds\int\rho(dx)\tilde\Psi^{1s}(x)](\xi^2)=I+II      $$
$$	I=\beta\int_0^\infty ds[\int_0^\infty dt\int\rho(dx)\,
	\nabla_x(\tilde\Psi^{1s}\circ f^t)\cdot F^2(x)]      $$
$$	II=-\beta\int_0^\infty ds\int\rho(dx)\,
	\nabla_x(D_\xi\tilde V(q(s))(\xi^2))\cdot F^1(x)      $$
$$	=\beta\int_0^\infty ds\int\rho(dx)
\nabla_x(\tilde G(f^sx)\cdot\int_0^sdt(T_{f^tx}f^{s-t})F^2(f^tx))\cdot F^1(x)$$
$$	=\beta\int_0^\infty ds\int_0^sdt
\int\rho(dx)\,\nabla_x(\tilde\Psi^{2(s-t)}\circ f^t)\cdot F^1(x)      $$
$$	=\beta\int_0^\infty ds\int_0^\infty dt\int\rho(dx)\,
	\nabla_x(\tilde\Psi^{2s}\circ f^t)\cdot F^1(x)      $$
where we have renamed $s$ the variable $s-t$ in the last line.  This proves (iii).
\medskip
	(iv) follows from (10).
\medskip
	(v) follows from (9), (iii) and (iv), where we take $\tilde\xi=\eta=\xi$.
\medskip\noindent
{\bf The gradient case.}
\medskip
	The situation where $\xi$ is a global gradient, {\it i.e.}, there is a potential function $V=V(q)$ such that $\xi(q)=-\nabla_qV$ is called {\it equilibrium} in the present context.  Let then 
$$	H(x)=h({p\cdot p\over 2m})e^{-\beta V(q)}      $$
The divergence of $HF_\xi$ is 
$$	\nabla_x\cdot(HF_\xi)
	=h'({p\cdot p\over2m})e^{-\beta V(q)}{p\over m}\cdot(\xi-\alpha p)
	+H[\nabla_p\cdot(-\alpha p)-\beta\nabla_qV\cdot{p\over m}]      $$
$$=H[-(N-1){p\cdot\xi(q)\over p\cdot p}+{\beta\over m}p\cdot\xi(q)]      $$
which vanishes if $h(p\cdot p/2m)=\delta(p\cdot p/2m-K)$ and $\beta=(N-1)/2K$.  Therefore the probability measure 
$$	\rho(dx)
=Z^{-1}\delta({p\cdot p\over 2m}-K)e^{-\beta V(q)}dp\,dq\eqno{(11)}      $$
(with normalizing factor $Z^{-1}$) is invariant under $(f^t)$ (see [6]) and is the SRB measure $\rho$ in the present case.  
\medskip
	Note that, using (11) and integrating by parts, we obtain 
$$	\int\rho(dx)\,\nabla_x\phi\cdot F^i(x)
	=Z^{-1}\int dx\,\nabla_x\phi\cdot\delta({p\cdot p\over 2m}-K)
	e^{-\beta V(q)}\pmatrix{\xi^i-\alpha^i p\cr0\cr}      $$
$$	=Z^{-1}\int dx\,\phi(x)\delta({p\cdot p\over 2m}-K)e^{-\beta V(q)}
	(-\nabla_x\cdot\pmatrix{\xi^i-\alpha^i p\cr0\cr})
	=\int\rho(dx)\,\phi(x)\Phi^i(x)\eqno{(12)}      $$
\medskip
	Define 
$$	Q(V)=\int e^{-\beta V(q)}dq      $$
Then the (configurational) Gibbs entropy associated with $\rho$ is 
$$  S(V)=-\int dq\,{e^{-\beta V(q)}\over Q(V)}\log{e^{-\beta V(q)}\over Q(V)} 
	=\rho(\beta V)+\log Q(V)      $$
If $V^1$ is a small perturbation of $V$ we find to first order in $V^1$
$$	S(V+V^1)-S(V)
=-[\rho((\beta V)(\beta V^1))-\rho(\beta V)\rho(\beta V^1)]\eqno{(13)}      $$
Using (8), (i) and (ii) of the above Proposition, and (12) we obtain 
$$	S(\xi\to\xi+\xi^1)
	=\beta\int_0^\infty ds\int\rho(dx)\nabla_xV(q(s))\cdot F^1(x)      $$
$$	=-\beta\int_0^\infty ds\int\rho(dx)\,
	V(q(s))(N-1){\nabla_qV^1\cdot p\over p\cdot p}      $$
$$	=-\beta^2\lim_{T\to\infty}\int_{-T}^0ds\int\rho(dx)
	V(q)\nabla_{q(s)}V^1\cdot{dq(s)\over ds}      $$
$$	=-\beta^2\lim_{T\to\infty}\int\rho(dx)
	V(q)[V^1(q)-V^1(q(-T))]=-\beta^2[\rho(VV^1)-\rho(V)\rho(V^1)]      $$
Therefore the standard estimate (13) from equilibrium statistical mechanics agrees with the ``nonequilibrium'' prediction based on (8).  
\medskip\noindent
{\bf Proposition.}
\medskip
	{\sl Assume that $\xi$ is a global gradient, then
\medskip
	(i) $\gamma_\xi(\xi^1,\xi^2)=\int_0^\infty ds\int_0^\infty dt\int\rho(dx)\,(\Phi^1\circ f^{s+t})\Phi^2(x)$
\medskip
	(ii) $R_\xi(\xi^1,\xi^2)=0$}
\medskip
	From (10) and (12), (i) directly follows.  We use now the involution $I:\pmatrix{p\cr q\cr}\mapsto\pmatrix{-p\cr q\cr}$, under which $\Phi^i$ is odd, time is reversed, and $\rho$ is invariant (``microscopic reversibility'').  Thus 
$$	\gamma_\xi(\xi^2,\xi^1)=\int_0^\infty ds\int_0^\infty dt
\int\rho(dx)\,\Phi^1(x)\Phi^2(f^{s+t}x)      $$
$$ =\int_0^\infty ds\int_0^\infty dt\int\rho(dx)\,\Phi^1(f^{-s-t}x)\Phi^2(x) $$
$$=\int_0^\infty ds\int_0^\infty dt\int\rho(dx)\,\Phi^1(f^{s+t}Ix)\Phi^2(Ix)$$
$$=\int_0^\infty ds\int_0^\infty dt\int\rho(dx)\,\Phi^1(f^{s+t}x)\Phi^2(x)
	=\gamma_\xi(\xi^1,\xi^2)      $$
and therefore $R_\xi=0$ by (v) of the previous proposition, proving (ii).  
\medskip\noindent
{\bf Second order formula.}
\medskip
	From the above considerations it follows that if $\xi$ is a gradient ($\xi=-\nabla V$), the entropy at temperature $\beta^{-1}$ can be written consistently to second order with respect to a (nongradient) perturbation $\xi^1$ of $\xi$ as 
$$	S(\xi+\xi^1)=S(\xi)+\gamma_\xi(\xi^1)
	+{1\over2}(D_\xi\gamma_\cdot(\xi^1))(\xi^1)\eqno{(14)}      $$
where $S(\xi)$ is the equilibrium entropy for $\xi$, and 
$$  \gamma_\xi(\xi^1)=\beta\int_0^\infty ds\int\rho(dx)V(q(s))\Phi^1(x)  $$
$$	{1\over2}(D_\xi\gamma_\cdot(\xi^1))(\xi^1)
	=\beta\int_0^\infty ds\int_0^\infty dt\int\rho(dx)\,
[{1\over2}\Phi^1(f^{s+t}x)\Phi^1(x)-(\Psi^{1s}\circ f^t)\Phi^1(x)]      $$
with $\Psi^{1s}=-\nabla_xV(q(s))\cdot F^1(x)$ and other notation explained earlier.  We have not studied $S(\xi+\xi^1)$ from the point of view of convexity.  
\medskip\noindent
{\bf Conclusion.}
\medskip
	In the present paper we have considered a classical system with isokinetic time evolution defined by (1) corresponding to a time independent force $\xi$ and temperature $\beta^{-1}$.  For such a system we have defined an entropy increment $S(\xi\to\xi+\delta\xi)$ corresponding to an increment $\delta\xi$ of the force (see (7), (8)).  Our definition agrees with the equilibrium statistical mechanics formula (for the Gibbs entropy) if $\xi$, $\delta\xi$ are gradient forces.  If $\xi$ is a gradient, but not necessarily $\delta\xi$, we can write 
$$	S(\xi+\delta\xi)=S(\xi)+S(\xi\to\xi+\delta\xi)      $$
where $S(\xi)$ is the equilibrium entropy and $S(\xi\to\xi+\delta\xi)$ is well defined by (14) to second order in $\delta\xi$.
\medskip\noindent
{\bf Acknowledgments.}
\medskip
	The present work was inspired by a number of discussions at Rutgers University with Giovanni Gallavotti, Joel Lebowitz, and Sheldon Goldstein.  Bill Hoover also kindly answered some email questions.  Let all of them be thanked.
\medskip\noindent
{\bf References.}
\medskip

[1] L. Andrey.  ``The rate of entropy change in non-Hamiltonian systems.''  Phys. Letters {\bf 11A},45-46(1985).

[2] N.I. Chernov, G.L. Eyink, J.L. Lebowitz, an Ya.G. Sinai.  ``Steady-state electrical conduction in the periodic Lorentz gas.''  Commun. Math. Phys. {\bf 154},569-601(1993).

[3] E.G.D. Cohen and L. Rondoni.  ``Note on phase space contraction and entropy production in thermostatted Hamiltonian systems.''  Chaos {\bf 8},357-365(1998).

[4] D. Dolgopyat.  ``On decay of correlations in Anosov flows.''  Ann. of Math. {\bf 147},357-390(1998).

[5] D. Dolgopyat.  ``Prevalence of rapid mixing for hyperbolic flows.''  Ergod. Th. and Dynam. Syst. {\bf 18},1097-1114(1998)

[6] D.J. Evans and G.P. Morriss.  {\it Statistical mechanics of nonequilibrium fluids.}  Academic Press, New York, 1990.

[7]  D.J. Evans and L. Rondoni.  ``Comments on the entropy of nonequilibrium steady states.''  J. Statist. Phys. {\bf 109},895-920(2002).  

[8] G. Gallavotti and E.G.D. Cohen.  ``Dynamical ensembles in stationary states.''  J. Statist. Phys. {\bf 80},931-970(1995).

[9] W.G. Hoover.  {\it Molecular dynamics.}  Lecture Notes in Physics {\bf 258}.  Springer, Heidelberg, 1986.

[10] W.G. Hoover.  {\it Time reversibility, computer simulation, and chaos.}  World Scientific, Singapore, 1999.

[11] J.L. Lebowitz.  ``Boltzmann's entropy and time's arrow.''  Physics Today {\bf 46}, No 9,32-38(1993).

[12] C. Liverani.  ``On contact Anosov flows.''  Preprint.

[13] D. Ruelle.  ``General linear response formula in statistical mechanics, and the fluctuation-dissipation theorem far from equilibrium.''  Phys. Letters {\bf A 245},220-224\ (1998).

[14] D. Ruelle.  ``Smooth dynamics and new theoretical ideas in nonequilibrium statistical mechanics.''  J. Statist. Phys. {\bf 95},393-468(1999).

\end